\documentclass[pre,preprint,notitlepage]{revtex4-1}%
\usepackage{amssymb}
\usepackage{amsfonts}
\usepackage{amsmath}
\usepackage{xcolor}
\usepackage{graphicx}%
\providecommand{\U}[1]{\protect\rule{.1in}{.1in}}
\providecommand{\U}[1]{\protect\rule{.1in}{.1in}}

\begin{document}
\title{The use of open boundaries in Mesoscopic Nucleation Theory}
\title{The use of open boundaries in stochastic hydrodynamic models of nucleation}
\author{James F. Lutsko}
\homepage{http://www.lutsko.com}
\email{jlutsko@ulb.be}
\affiliation{Center for Nonlinear Phenomena and Complex Systems CP 231 and BLU-ULB Space Research Center, Universit\'{e}
  Libre de Bruxelles, Blvd. du Triomphe, 1050 Brussels, Belgium}

\begin{abstract}
  Stochastic hydrodynamics is a central tool in the  study of first order phase transitions at a fundamental level. Combined with sophisticated free energy models,
  e.g. as developed in classical Density Functional Theory, complex processes such as crystallization can be modeled and information such as
  free energy barriers, nucleation pathways and the unstable eigenvector and eigenvalues determined. The latter are particularly interesting as they play key roles in defining
  the natural (unbiased) order parameter and the nucleation rate respectively. As is often the case, computational realities restrict the size of system that
  can be modeled and this makes it difficult to achieve experimental conditions for which the volume is effectively infinite. In this paper, the use of open boundary conditions is
  discussed. By using an open system, the calculations become much closer to experimental conditions however, the introduction of open boundary conditions raises a number of questions
  concerning the stochastic model such as whether the fluctuation-dissipation relation is preserved and whether stationary points on the free energy surface remain stationary points of the dynamics. These
  questions are addressed here and illustrated with calculations on realistic systems. 
\end{abstract}

\maketitle

\section{Introduction}

Stochastic models are an important tool used to bridge the gap between exact,
microscopic dynamics and macroscopic, observable phenomenon\cite{Gardiner}.
Historically, they have often arisen based on physical heuristics - such as
the Einstein-Smolochowski description of diffusion and Landau's fluctuating
hydrodynamics\cite{Landau,Lifshitz}- but they can also be founded in more systematic developments
based, e.g. on projection operator techniques\cite{Grabert,Zubarev,Duran-Olivencia}. The stochastic
component that distinguishes them captures such effects as thermal
fluctuations and interactions with small particles (e.g. a bath) which play a
crucial role in diverse phenomena such as protein folding\cite{ProteinFolding}, the
wetting transition\cite{Wetting, Evans_Wetting}, the Cassie-Baxter to Wenzel transition associated with the filling of pores\cite{superhydrophobic} and first order phase
transitions\cite{Kashchiev}. Recently, one such model consisting of a
combination of fluctuating hydrodynamics and classical Density\ Functional
Theory\ (cDFT)\cite{Evans79,LutskoAdvChemPhysDFT}, called Mesoscopic
Nucleation Theory (MNT)\cite{Lutsko_JCP_2011_Com, Lutsko_JCP_2012_1,
Lutsko_Schoonen_JCP}, has been used to study first order phase transitions and
particularly the complex phenomenon of crystallization
\cite{HCF,Lutsko_Schoonen_JCP}. Fundamental work on the microscopic
foundations of this model\cite{Duran} combined with the analysis of its
relation to the heuristic, macroscopic Classical Nucleation Theory
(CNT)\cite{Lutsko_JCP_2012_1} - as well as to intermediate approaches such as
Phase Field Theory\cite{Lowen} - has laid the foundations for a statistical
mechanical description of the dynamics of first order nucleation in general and of
crystallization in particular.

Fluctuating hydrodynamics describes the time evolution of the local densities
of the conserved quantities mass, momentum and energy driven by gradients of
mass, pressure, temperature and by imposed external forces plus, of
course, thermal fluctuations\cite{Landau,Lifshitz}. As is the case with any coarse
grained model, various quantities must be approximated, even if exact
microscopic expressions for them exist. A key example in the application of
fluctuating hydrodynamics to phase transitions is the local pressure. In MNT,
this is related to a free energy functional which for which models developed
in cDFT are used. The cDFT free energies are functionals of the local density
and are based on rigorous results from equilibrium statistical mechanics. They
accurately describe the spatial structure of inhomogeneous systems at all
length scales and are thus suitable for studying crystallization in which
molecular-scale inhomogeneities in the density are fundamental\cite{LutskoLam}%
. An important application for which it is possible to simplify the problem is
that of nucleation of colloids (and macromolecules) from solution in which
case it is reasonable to work in the over-damped regime. This has the
advantage that the hydrodynamic description reduces to a single equation for
the local density and the resulting model is recognized as a version of the
Dean-Kawasaki model introduced in other contexts\cite{Dean,Kawasaki1998,Chavanis_Deriv} and
also known as stochastic Dynamical Density Functional Theory\cite{Illien}.

When a dense phase, such as a dense droplet or a crystal, nucleations from
solution in a closed system the concentration of the solution necessarily
decreases which in turn changes the characteristics (e.g. free energy barrier
and nucleation rate) of the process. However, in many applications, the
volumes are macroscopic and the rate of nucleation very small so that for all
practical purposes, the concentration remains effectively constant. In
contrast, numerical applications of MNT, as in simulations, are generally
restricted to much smaller volumes and so any nucleation has a strong effect
on the concentration of a closed system. For this reason, to achieve
correspondence with the common experimental situation it is more practical to
consider open systems in which mass is not conserved. Since MNT is based on
hydrodynamics, in which mass is concerved, the most natural way to do this is
via the boundary conditions imposed on the system. The present work is an
exploration of the the formal and technical consequences of applying such open
boundary conditions to Dean-Kawasaki type models. Three main formal issues
will be addressed. First, since these models usually involve state-dependent
noise, their interpretation in terms of stochastic processes is a technical,
but not trivial point and it will be seen that open boundaries complicates the
discussion. This is not only a mathematical issue but in fact is required so
that the free energy functional used is actually the free energy of the system
when it is in equilibrium.  The second issue to be considered is particularly
motivated by the application to nucleation and involves compatability of the
dynamics with the usual predictions of cDFT. Classical DFT allows one to
determine the critical cluster and properties, which plays a central role in
understanding nucleation and it is reasonable to demand the the dynamic model
respects these results:\ that is, to demand that minima and saddle points on
the free energy surface are stationary states of the dynamical model. As is
explained below, this is again not just a formal requirement but allows for
the use of very convenient techniques in actual applications. The third and
final issue is the effect of the boundaries on the calculation of unstable
modes of critical clusters, which is another key quantity in the description
of nucleation.

The analysis presented here is phrased in terms of the specifics of MNT
although one expects that the conclusions can easily be extrapolated to other
applications. For this reason, the next Section gives a brief overview of the
cDFT that is the source of the free energy functionals used in the stochastic
model, its particular connection to nucleation and how numeric calculations
are formulated. The third section addresses the dynamics including the
important constraint of the fluctuation-dissipation relation, and the way in
which fixed boundaries naturally lead to the transition from a canonical free
energy surface to a grand-canonical surface and the important role of the
choice of discretization scheme on the formal properties of the discretized
model. Section IV presents some illustrative calculations showing strengths
and weaknesses of the various discretization schemes. In the final, concluding
Section, the practical take-aways of the analysis are summarized.

\section{Classical DFT}

\subsection{cDFT formalism}

The foundations of cDFT is closely related to that of the more well-known
quantum DFT but its practical development is very different\cite{Evans79,LutskoAdvChemPhysDFT}. Like qDFT, it is
formulated in terms of a local density $\rho\left(  \mathbf{r}\right)  $,
although in cDFT is the mass or number density whereas in qDFT it is the local
electron density. Similarly, cDFT is based on the proof of the existence of a
functional of the local density $\Omega_{\mu TV}\left[  n;\varphi\right]  $
which, as indicated, depends on the chemical potential $\mu$, the temperature
$T$,the volume $V$ and any external field $\varphi\left(  \mathbf{r}\right)  $
acting on the system and which has the property that it is minimized by the
equilibrium density distribution $\rho^{\text{eq}}\left(  \mathbf{r}\right)  $.
Furthermore, the value of the functional evaluated at the minimum is the
grand-canonical free energy, $\Omega_{\mu TV}=\Omega_{\mu TV}\left[
\rho^{\text{eq}};\varphi\right]  $. If more than one density gives the same minimum, then
multiple coexisting states are possible. Thus, the equilibrium density can be
found by minimizing the functional. The functional $\Omega$ has the structure%
\begin{equation}
\Omega_{\mu TV}\left[\rho;\varphi\right]  =F_{TV}\left[\rho\right]  +\int
_{V}\left(  \varphi\left(  \mathbf{r}\right)  -\mu\right)  \rho\left(
\mathbf{r}\right)  d\mathbf{r}%
\end{equation}
where the field-independent functional $F\left[  n\right]  $, usually called
the Helmholtz functional, is not known except for some special cases\cite{LutskoAdvChemPhysDFT,esFMT,LutskoCanonical}.

To determine the equilibrium state(s), one can take the functional derivative
with respect to density and set it equal to zero resulting in the
Euler-Lagrange equation%
\begin{equation}
\frac{\delta F_{TV}\left[  n\right]  }{\delta n\left(  \mathbf{r}\right)
}=\mu-\varphi\left(  \mathbf{r}\right)  . \label{EL}%
\end{equation}
Note however that solutions to this equation are not just quasi-stable minima
on the free energy surface but more generally, stationary points including
unstable saddle points. In the absence of external fields and in the
thermodynamic limit of very large volumes, the minima are typically either
homogeneous, $\rho^{\text{eq}}\left(  \mathbf{r}\right)  =\overline{\rho}$, corresponding
to vapor and liquid phases, or crystalline solids. Often, three or more such
minima exist with the thermodynamic equilibrium state being the absolute
minimum. Since the minima are stable, they divide density space into regions
attracted to each minimum with the border being any two such regions being a
separatrix. The lowest point on each separatrix is a saddle point or,
physically, a critical cluster and these also, since they are stationary
points, solve the Euler-Lagrange equation.

\subsection{cDFT and Nucleation:\ finding critical clusters}

Critical clusters are the key objects in studying nucleation so it is
important to be able to find them. A variety of methods are available for the
general problem of locating saddle points on a high-dimensional energy
surface\cite{Wales} and here we only mention two methods which are
particularly useful for the nucleation problem. First, consider the nucleation
of a condensed phase from a low density (vapor) mother phase in a system with
a fixed number of particles. In this case, as clusters grow, the density of
the mother phase outside the cluster decreases which causes an increase in
free energy of the vapor background. At some point, a critical size is reached
and, just as for the grand-canonical case, if the cluster grows slightly
larger than the critical size it becomes super-critical and the free energy of
the whole system is lowered. However, since the density of the background is
decreasing as it grows, at some point when the cluster becomes sufficiently
large, a stable equilibrium is reached - the limit would have to be sometime
before the background density reaches zero. So for the constant-mass system,
there is a stable cluster larger than the critical cluster. Now, if the
constraint of constant mass is removed - in other words, if we open the system
- this stable cluster becomes unstable because it can grow without decreasing
the background density. Thus,  the stable cluster for the closed system is a
saddle point - i.e. a critical cluster - for the open system. This intuition
can easily be formalized\cite{LutskoLam}. One can minimize the free energy
functional at constant particle number via the introduction of a Lagrange
multiplier $\lambda$, one minimizes the Lagrangian
\begin{equation}
L\left[\rho\right]  =\Omega_{\mu^{\prime}TV}\left[\rho;\varphi\right]
+\lambda\left(  N-\int_{V} \rho\left(  \mathbf{r}\right)  d\mathbf{r}\right)  ,
\end{equation}
where $\mu^{\prime}$ can be anything, giving the Euler-Lagrange equations
\begin{align}
\frac{\delta F_{TV}\left[\rho\right]  }{\delta \rho\left(  \mathbf{r}\right)  } &
=\mu^{\prime}-\varphi\left(  \mathbf{r}\right)  -\lambda\label{EL1}\\
\int_{V}\rho\left(  \mathbf{r}\right)  d\mathbf{r} &  =N\nonumber
\end{align}
to be solved simultaneously for $\lambda$ and $\rho\left(  \mathbf{r}\right)  $.
If a cluster-like solution is found, and in practice this is generally the
case if starting from any cluster-like initial density, then comparing the
first of Eq.(\ref{EL1}) to the open-system Euler-Lagrange equation (\ref{EL}),
one sees that the solution to the former is also a solution to the latter for
the chemical potential $\mu=\mu^{\prime}-\lambda$. This method has the
advantage that the same techniques (and code) used to find stable solutions to
the open-system Euler-Lagrange equation can be re-used to find critical
clusters. The disadvantage is that one does not control any properties - such
as the chemical potential, background density or critical size - directly.

A second method, also used in the present work, is to minimize the magnitude
of the forces. Specifically, instead of minimizing $L\left[  n\right]  $ for a
closed system, one can minimize the squared forces by solving
\begin{equation}
0=\frac{\delta}{\delta \rho\left(  \mathbf{r}\right)  }\int_{V}\left(
\frac{\delta L\left[  \rho\right]  }{\delta \rho\left(  \mathbf{r}^{\prime}\right)
}\right)  ^{2}d\mathbf{r}^{\prime}=2\int_{V}\left(  \frac{\delta L\left[
n\right]  }{\delta \rho\left(  \mathbf{r}^{\prime}\right)  }\right)  \left(
\frac{\delta^{2}L\left[  \rho\right]  }{\delta \rho\left(  \mathbf{r}^{\prime
}\right)  \delta \rho\left(  \mathbf{r}\right)  }\right)  d\mathbf{r}^{\prime}%
\end{equation}
provided that the right hand side can be conveniantly evaluated. This turns
out to be the case for the cDFT functionals used in MNT, see below, so that
this is a practical alternative to the previous method which has the advantage
that it can be directly performed on an open system. While any stationary
density distribution is clearly a minimum of the functional, the disadvantage
is that one must start with some sort of "reasonable" guess or risk falling
down to a trivial solution. For the example of droplet nucleation discussed
below, it is the case that even quite crude capillary approximations to a
critical cluster are generally good enough starting points.  

\subsection{Discretized model}

The equations will be discretized on a cubic grid with lattice points
$\mathbf{r}_{\mathbf{I}}=\Delta\mathbf{I}_{a}\hat{\mathbf{e}}_{a}$ where
$\Delta$ is the lattice constant, $\hat{\mathbf{e}}_{a}$ for $a=x,y,z$ are the
cartesian directions and the superindex $\mathbf{I}=(I_{x},I_{y},I_{z})$ with
$0\leq I_{a}<N_{a}-1$. In the following, these are combined so that e.g.
$\mathbf{I}\pm\hat{\mathbf{e}}_{x}\equiv(I_{x}\pm1,I_{y},I_{z})$. The density
at lattice point $\mathbf{r}_{\mathbf{I}}$ is written as $\rho_{\mathbf{I}}\equiv
\rho(\mathbf{r}_{\mathbf{I}})$. There is often need in the following to refer to
the collection of all of the discrete density variables so these will be
denoted collectively as $\rho$. Periodic boundaries are assummed in all
calculations of the free energy functionals with periodicity \ $\mathbf{N}%
=\left(  N_{x},N_{y},N_{z}\right)  $ so that $\rho_{\mathbf{I}}=\rho_{\mathbf{I+}%
N_{a}\widehat{\mathbf{e}}_{a}}$ for $a\in(x,y,z)$. This means that, with no
other constraints, there are a total of $N_{x}\times N_{y}\times N_{z}$
independent values of the density and all sums are over all of these
components. The free energy functional then becomes a function of the
variables $\rho_{\mathbf{I}}$, $\Omega_{\mu TV}\left[\rho;\varphi\right]
\rightarrow\Omega_{\mu TV}\left(\rho;\varphi\right)  $, e.g.%
\[
\int_{V} \rho\left(  \mathbf{r}\right)  \varphi\left(  \mathbf{r}\right)
d\mathbf{r}\rightarrow\sum_{\mathbf{I}\in\mathcal{V}}\rho_{\mathbf{I}}%
\varphi_{\mathbf{I}}\Delta^{3}%
\]
where $\mathcal{V}$ is the set of all valid indices, $\left(  0,0,0\right)  $
to $\left(  N_{x}-1,N_{y}-1,N_{z}-1\right)  $, and functional derivatives
become ordinary derivatives%
\begin{equation}
\frac{\delta}{\delta \rho\left(  \mathbf{r}\right)  }\rightarrow\frac{1}%
{\Delta^{3}}\frac{\partial}{\partial \rho_{\mathbf{I}}}.
\end{equation}

Details of the specific cDFT model used in this work have been given
elsewhere\cite{Schoonen}. In brief, taking the simplest case as illustration, the Helmholtz
functional for a single-species system of molecules interacting via the pair
potential $v\left(  \mathbf{r}_{1}-\mathbf{r}_{2}\right)  $, is written as the
sum of three contributions, $F_{TV}\left[\rho\right]  =F_{TV}^{\left(
\text{id}\right)  }\left[\rho\right]  +F_{TV}^{\left(  \text{HS}\right)
}\left[\rho\right]  +F_{TV}^{\left(  \text{mf}\right)  }\left[\rho\right]  $.
The ideal gas contribution is
\begin{equation}
F_{TV}^{\left(  \text{id}\right)  }\left[\rho\right]  =\sum_{\mathbf{I}%
\in\mathcal{V}}\left(\rho_{\mathbf{I}}\ln\left(\rho_{\mathbf{I}}\Lambda
^{3}\right)  -\rho_{\mathbf{I}}\right)  \Delta^{3},
\end{equation}
where $\Lambda$ is the thermal wavelength and the hard-sphere contribution is%
\begin{equation}
F_{TV}^{\left(  \text{HS}\right)  }\left[\rho\right]  =\sum_{\mathbf{I}%
\in\mathcal{V}}\Phi\left(  \eta^{\left(  1\right)  },\eta^{\left(  2\right)
}...\right)  \Delta^{3},\;\eta_{\mathbf{I}}^{\left(  \alpha\right)  }\left(
\rho\right)  =\sum_{\mathbf{J}\in\mathcal{V}}w_{\mathbf{I-J}}^{\left(
\alpha\right)  }\rho_{\mathbf{J}}\Delta^{3}%
\end{equation}
where $\Phi$ is a function of the so-called fundamental measures, $\eta_{\mathbf{I}%
}^{\left(  \alpha\right)  }$, which in turn are expressed in terms of
convolutions of the density with weights $w_{\mathbf{I-J}}^{\left(
\alpha\right)  }$\cite{Rosenfeld1,EvansRoth,LutskoAdvChemPhysDFT} (the present calculations use the explicitly stable fundamental measure theory\cite{esFMT}). The final contribution is a mean-field term
\begin{equation}
F_{TV}^{\left(  \text{mf}\right)  }\left[\rho\right]  =\frac{1}{2}%
\sum_{\mathbf{I,J}\in\mathcal{V}}\rho_{\mathbf{I}}\rho_{\mathbf{J}}v_{\mathbf{I-J}%
}^{\left(  \text{att}\right)  }\Delta^{6}%
\end{equation}
where $v_{\mathbf{I-J}}^{\left(  \text{att}\right)  }$ is the discretized form
of the attractive part of the pair potential describing molecular
interactions. Note that the mean field term, as well as the fundamental
measures, are convolutions and so can be efficiently evaluated with fast
Fourier transforms (FFTs). The derivative of the free energy with respect to
the density is
\begin{equation}
\frac{\partial F\left(\rho\right)  }{\partial \rho_{\mathbf{I}}}=\ln\left(
\rho_{\mathbf{I}}\Lambda^{3}\right)  \Delta^{3}+\sum_{\mathbf{J}\in\mathcal{V}%
}\sum_{\alpha}\frac{\partial\Phi}{\partial\eta_{\mathbf{J}}^{\left(
\alpha\right)  }}w_{\mathbf{J-I}}^{\left(  \alpha\right)  }\Delta^{6}%
+\sum_{\mathbf{J}\in\mathcal{V}}\rho_{\mathbf{J}}v_{\mathbf{I-J}}^{\left(
\text{att}\right)  }\Delta^{6}%
\end{equation}
and%
\begin{equation}
\frac{\partial F\left(\rho\right)  }{\partial \rho_{\mathbf{I}}\partial
\rho_{\mathbf{J}}}=\frac{\Delta^{3}}{\rho_{\mathbf{I}}}+\sum_{\mathbf{K,L}%
\in\mathcal{V}}\sum_{\alpha,\gamma}\frac{\partial^{2}\Phi}{\partial
\eta_{\mathbf{K}}^{\left(  \alpha\right)  }\partial\eta_{\mathbf{L}}^{\left(
\gamma\right)  }}w_{\mathbf{K-I}}^{\left(  \alpha\right)  }w_{\mathbf{L-J}%
}^{\left(  \alpha\right)  }\Delta^{9}+v_{\mathbf{I-J}}^{\left(  \text{att}%
\right)  }\Delta^{6}%
\end{equation}
which can, again, be efficiently evaluated with discrete FFTs.

\section{Dynamical Formalism}

\subsection{Stochastic dynamics and fluctuation-dissipation relation}

The stochastic differential equation governing the dynamics of over-damped MNT
is
\begin{equation}
\label{DK}\frac{\partial}{\partial t}\rho_{t}(\mathbf{r})=D\nabla\cdot\rho
_{t}(\mathbf{r})\nabla\frac{\delta\beta F[n]}{\delta n(\mathbf{r}%
)}|_{n(\mathbf{r})=\rho_{t}(\mathbf{r})}+\nabla\cdot\sqrt{2D\rho
_{t}(\mathbf{r})}\;\;\mathbf{\xi}(\mathbf{r},t)
\end{equation}
where $D$ is the tracer diffusion constant, the square brackets indicate a
functional dependence and $\delta$ indicates a functional derivative. The free
energy $F[\rho]$ is a functional of the local density and $\mathbf{\xi
}(\mathbf{r},t)$ represents gaussian white noise, so $\left\langle
\mathbf{\xi}(\mathbf{r},t)\mathbf{\xi}(\mathbf{r}^{\prime},t^{\prime
})\right\rangle =\delta(\mathbf{r}-\mathbf{r}^{\prime})\delta(t-t^{\prime})$.
This can be expressed in a more abstract, but useful, form as
\begin{equation}
\frac{\partial}{\partial t}\rho_{t}(\mathbf{r})=-D\int_{V}g(\mathbf{r}%
,\mathbf{r}^{\prime};[\rho_{t}])\frac{\delta\beta F[n]}{\delta
n(\mathbf{r^{\prime}})}|_{n(\mathbf{r^{\prime}})=\rho_{t}(\mathbf{r^{\prime}%
})}d\mathbf{r}^{\prime}+\sqrt{D}\int_{V}\sum_{a\in\{x,y,z\}}q_{a}%
(\mathbf{r},\mathbf{r}^{\prime};[\rho_{t}])\mathbf{\xi}_{a}(\mathbf{r^{\prime
}},t)d\mathbf{r}^{\prime}%
\end{equation}
where $V$ is the integration volume and
\begin{align}
g(\mathbf{r},\mathbf{r}^{\prime};[\rho_{t}])  &  =-\nabla\cdot\rho
_{t}(\mathbf{r})\nabla\delta(\mathbf{r}-\mathbf{r}^{\prime})\\
\mathbf{q}(\mathbf{r},\mathbf{r}^{\prime};[\rho_{t}])  &  =\boldsymbol{\nabla
}\sqrt{2\rho_{t}(\mathbf{r})}\delta(\mathbf{r}-\mathbf{r}^{\prime}).\nonumber
\end{align}
It is shown in the Appendix that, for a rectilinear system with
periodic boundaries, these operators obey a fluctuation-dissipation relation,
\begin{equation}
\int_{V}\mathbf{q}(\mathbf{r},\mathbf{r}^{\prime};[\rho])\cdot
\mathbf{q}(\mathbf{r^{\prime\prime}},\mathbf{r}^{\prime};[\rho%
])d\mathbf{r}^{\prime}=2g(\mathbf{r},\mathbf{r}^{\prime\prime};[\rho]),
\end{equation}
which assures that the operator $g(\mathbf{r},\mathbf{r}^{\prime\prime};[\rho])$ is positive semi-definite.

A stationary state of the deterministic part of the dynamics is a density
distribution, $\rho^{\text{stat}}$ for which $\nabla\cdot\rho(\mathbf{r})
\nabla\frac{\delta F[\rho]}{\delta \rho(\mathbf{r})} |_{\rho^{\text{stat}}(\mathbf{r})}=0$ . In
practice, the most important such states for a system with periodic boundaries
are those for which $\frac{\delta F[\rho]}{\delta \rho(\mathbf{r})} |_{\rho^{\text{stat}}(\mathbf{r}%
)}=\lambda$ where $\lambda$ is a constant. It proves convenient - indeed,
crucial - for the case of open systems to modify the force as $F[\rho]
\rightarrow\Omega[\rho;\lambda] \equiv F[\rho]-\lambda N[\rho]$ where the total number
of particles in the \textbf{computational volume} is $N[\rho] \equiv\int_{V}
\rho(\mathbf{r}) d\mathbf{r}$. This clearly makes no difference for periodic
systems since translational invariance means that the gradients give zero
acting on the constant $\lambda$ . However, for open systems in which we
freeze the density on the boundaries, this term will be critical in preserving
the usual stationary states, as will be made evident below.

\subsection{Discretized equations}

The fact that the right hand side of Eq.(\ref{DK}) is a total gradient implies
that mass is conserved when periodic boundaries are used and so we refer to
this as a \textbf{closed system}. To model an \textbf{open system}, we will
fix the densities on the border to a given value, $\rho_{B}$, in which case
mass is not conserved. In the following, we define $\mathcal{V}$ to be the set
of all lattice points, $\partial\mathcal{V}$ to be the lattice points on the
boundary and so the interior, non-boundary, points are $\mathcal{V}%
-\partial\mathcal{V}$. Finally, we let $\mathcal{D}$ be the set of
\emph{dynamical} points meaning those points for which the density is a
dynamical variable that can change in time. So, $\mathcal{D}=\mathcal{V}$ for
periodic boundaries (closed systems) and $\mathcal{D}=\mathcal{V}%
-\partial\mathcal{V}$ for fixed boundaries (open systems).

Differential operators, like $\nabla$ will be discretized using either forward
or central differences. The operator $\nabla\cdot\rho_{t}(\mathbf{R}%
)\nabla\rightarrow g_{\mathbf{IJ}}(\rho_{t})$ where the form of the matrix
$g_{\mathbf{IJ}}(\rho_{t})$ will be given explicitly below. The discretized
dynamical equations can then be written as
\begin{equation}
\frac{\partial\rho_{t\mathbf{I}}}{\partial t}=-D\sum_{\mathbf{J}\in
\mathcal{D}}g_{\mathbf{IJ}}\left(  \rho_{t}\right)  \frac{\partial\beta
F\left(  \rho_{t}\right)  }{\partial\Delta^{3}\rho_{t\mathbf{J}}}+\sqrt
{\frac{2D}{\Delta^{3}}}\sum_{a\in(x,y,z)}\sum_{\mathbf{J}\in\mathcal{V}%
}q_{\mathbf{I}\mathbf{J}a}\left(  \rho_{t}\right)  \xi_{t\mathbf{J}%
a},\;\mathbf{I}\in\mathcal{D},\;\left\langle \xi_{t\mathbf{I}a}\xi_{t^{\prime
}\mathbf{J}b}\right\rangle =\delta_{ab}\delta_{\mathbf{IJ}}\delta\left(
t-t^{\prime}\right)  \label{A}%
\end{equation}
or
\begin{equation}
\frac{\partial\rho_{t\mathbf{I}}}{\partial t}=-D\sum_{\mathbf{J}\in
\mathcal{D}}g_{\mathbf{IJ}}\left(  \rho_{t}\right)  \frac{1}{\Delta^{3}}\beta
F_{\mathbf{J}}\left(  \rho_{t};\lambda\right)  +\sqrt{\frac{2D}{\Delta^{3}}%
}\sum_{a\in(x,y,z)}\sum_{\mathbf{J}\in\mathcal{V}}q_{\mathbf{I}\mathbf{J}%
a}\left(  \rho_{t}\right)  \xi_{t\mathbf{J}a},\;\mathbf{I}\in\mathcal{D}%
\end{equation}
where here and in the following the compact notation
\begin{align}
\beta F_{\mathbf{J}}\left(  \rho\right)   &  \equiv\frac{\partial\beta
F\left(  \rho\right)  }{\partial\rho_{\mathbf{J}}},\;\;\mathbf{J}%
\in\mathcal{D}\\
\beta F_{\mathbf{IJ}}\left(  \rho\right)   &  \equiv\frac{\partial^{2}\beta
F\left(  \rho\right)  }{\partial\rho_{\mathbf{I}}\partial\rho_{\mathbf{J}}%
},\;\;\mathbf{J}\in\mathcal{D}\nonumber
\end{align}
is used. The sums over the noise terms deliberately include all points, and
not just the dynamical points, for reasons explained below. The
fluctuation-dissipation relation becomes
\begin{equation}
g_{\mathbf{IJ}}\left(  \rho\right)  =\sum_{a\in\{x,y,z\}}\sum_{\mathbf{K}%
\in\mathcal{V}}q_{\mathbf{IK}a}q_{\mathbf{JK}a}%
\end{equation}
and demanding that any discretization of the model satisfy this relation is crucial since it assures that
the matrix $g(\rho)$ is positive semidefinite and this in turn is necessary so as not to introduce spurious
unstable modes into the dynamics.

We recall that in the case of state-dependent noise, as is the case here, the
stochastic differential equation must be supplemented with an
\emph{interpretation}. In the case of a discretized time variable, the
interpretation specifies whether the noise amplitude is to be evaluated at the
beginning of the time step (the Ito interpretation), at the middle of the time
step (the Stratonovich interpretation) or the end of the time-step (the
anti-Ito interpretation). The latter two result in semi-implicit and implicit
discretization schemes, respectively, and - unlike the case of the
deterministic term - do not give identical results in the limit that the
time-step goes to zero. A practical consequence is the form of the
Fokker-Planck equation for the probability that the system has state values
$\rho$ at time $t$, $P(\rho,t)$ which, in the present case,
is
\begin{equation}
\frac{\partial}{\partial t}P_{t}\left(  \rho\right)  =-\sum_{\mathbf{I,J}%
\in\mathcal{D}}\frac{\partial}{\partial\rho_{\mathbf{I}}}\left(
Dg_{\mathbf{IJ}}\left(  \rho\right)  \beta F_{\mathbf{J}}\left(  \rho\right)
+\sqrt{\frac{2D}{\Delta^{3}}}\left\{
\begin{array}
[c]{c}%
\frac{\partial}{\partial\rho_{\mathbf{J}}}g_{\mathbf{IJ}}\left(  \rho\right)
\\
\sum_{a}\sum_{\mathbf{K}\in V}q_{\mathbf{I}\mathbf{K}a}\left(  \rho\right)
\frac{\partial}{\partial\rho_{\mathbf{J}}}q_{\mathbf{J}\mathbf{K}a}\left(
\rho\right)  \\
g_{\mathbf{IJ}}\left(  \rho\right)  \frac{\partial}{\partial\rho_{\mathbf{J}}}%
\end{array}
\right\}  \right)  P_{t}\left(  \rho\right)  \label{FP}%
\end{equation}
for the three cases (going from top to bottom), respectively. The Ito and
Stratonovich interpretations are equivalent if and only if
\begin{equation}
\sum_{\mathbf{J}\in\mathcal{D}}\sum_{a\in\{x,y,z\}}\sum_{\mathbf{K}%
\in\mathcal{V}}\left(  \frac{\partial}{\partial\rho_{\mathbf{J}}}%
q_{\mathbf{I}\mathbf{K}a}\left(  \rho\right)  \right)  q_{\mathbf{J}%
\mathbf{K}a}\left(  \rho\right)  =0
\end{equation}
and the Stratonovich and anti-Ito are the same if
\begin{equation}
\sum_{\mathbf{J}\in\mathcal{D}}\sum_{a\in\{x,y,z\}}\sum_{\mathbf{K}%
\in\mathcal{V}}q_{\mathbf{I}\mathbf{K}a}\left(  \rho\right)  \left(
\frac{\partial}{\partial\rho_{\mathbf{J}}}q_{\mathbf{J}\mathbf{K}a}\left(
\rho\right)  \right)  =0
\end{equation}
In a well-known contribution to the theory of fluctuating hydrodynamics, from
which this over-damped dynamics is derived, it was shown that for closed
systems, discretization with centered finite differences resulted in
Ito-Stratonovich equivalence\cite{Saarloos}. One goal below is to explore this
for an open system. Finally, note from Eq. (\ref{FP}) that for the anti-Ito
interpretation, stationary distributions (i.e. setting $\frac{\partial
}{\partial t}P_{t}\left(  \rho\right)  =0$) include
\begin{equation}
P\left(  \rho\right)  =Ae^{-\beta F\left(  \rho\right)  },
\end{equation}
where $A$ is any constant, as expected. This is a strong motivation for
preferring models that are the same in all statistical interpretations. 

\subsection{Central differences}

Writing the gradient operator as $\nabla_{a}\rightarrow\frac{1}{2\Delta
}\left(  \delta_{\mathbf{K}}^{\mathbf{I+}\hat{\mathbf{e}}_{\left(  a\right)
}}-\delta_{\mathbf{K}}^{\mathbf{I-}\hat{\mathbf{e}}_{\left(  a\right)  }%
}\right)  $ results in the discretization
\begin{equation}
q_{\mathbf{IK}a}(\rho)=\frac{1}{2\Delta}\left(  \delta_{\mathbf{I+}%
\hat{\mathbf{e}}_{\left(  a\right)  }\mathbf{K}}-\delta_{\mathbf{I-}%
\hat{\mathbf{e}}_{\left(  a\right)  }\mathbf{K}}\right)  \sqrt{\rho
_{\mathbf{K}}},\;\mathbf{I}\in\mathcal{D},\mathbf{K}\in\mathcal{V}%
\end{equation}
giving, via the fluctuation-dissipation relation%
\begin{equation}
g^{\mathbf{IJ}}=\sum_{a\in\{x,y,z\}}\sum_{\mathbf{K}\in\mathcal{V}%
}q_{\mathbf{IK}a}(\rho)q_{\mathbf{JK}a}(\rho)=-\frac{1}{4\Delta^{2}}\sum
_{a\in\{x,y,z\}}\left\{  \rho_{\mathbf{I+}\hat{\mathbf{e}}_{\left(  a\right)
}}\left(  \delta_{\mathbf{I+}2\hat{\mathbf{e}}_{\left(  a\right)  }\mathbf{J}%
}-\delta_{\mathbf{I}\mathbf{J}}\right)  -\rho_{\mathbf{I-}\hat{\mathbf{e}%
}_{\left(  a\right)  }}\left(  \delta_{\mathbf{I}\mathbf{J}}-\delta
_{\mathbf{I-}2\hat{\mathbf{e}}_{\left(  a\right)  }\mathbf{J}}\right)
\right\}  ,\;\mathbf{I,J}\in\mathcal{D}.
\end{equation}
Using the reverse mapping $f_{\mathbf{I\pm}2\hat{\mathbf{e}}_{\left(
a\right)  }}\rightarrow f\left(  \mathbf{r}_{\mathbf{I}}\pm2\Delta
\hat{\mathbf{e}}_{\left(  a\right)  }\right)  $, etc. , it is straightforward
to show that away from the boundaries (in the sense that $\mathbf{I}%
,\mathbf{I\pm}2\hat{\mathbf{e}}_{\left(  a\right)  }\in\mathcal{D}$),
\begin{equation}
\lim_{\Delta\rightarrow0}\left(  \sum_{\mathbf{J}\in\mathcal{D}}%
g_{\mathbf{IJ}}f_{\mathbf{J}}\right)  =-\left(  \nabla\cdot\rho\left(
\mathbf{r}\right)  \left(  \nabla f\left(  \mathbf{r}\right)  \right)
\right)  _{\mathbf{r}_{\mathbf{I}}}%
\end{equation}
as required.

For an arbitrary vector $f_{\mathbf{I}}$,
\begin{equation}
v_{\mathbf{I}}\equiv\sum_{\mathbf{J}\in\mathcal{D}}g_{\mathbf{IJ}%
}f_{\mathbf{J}}=-\frac{1}{4\Delta^{2}}\sum_{a\in\{x,y,z\}}\left\{
\rho_{\mathbf{I+}\hat{\mathbf{e}}_{\left(  a\right)  }}\left(  \widetilde
{f}_{\mathbf{I+}2\hat{\mathbf{e}}_{\left(  a\right)  }}-\widetilde
{f}_{\mathbf{I}}\right)  -\rho_{\mathbf{I-}\hat{\mathbf{e}}_{\left(  a\right)
}}\left(  \widetilde{f}_{\mathbf{I}}-\widetilde{f}_{\mathbf{I-}2\hat
{\mathbf{e}}_{\left(  a\right)  }}\right)  \right\}  ,\;\mathbf{I}%
\in\mathcal{D}\label{I1}%
\end{equation}
where we define  $\widetilde{f}_{\mathbf{I}}=f_{\mathbf{I}}$ for
$\mathbf{I}\in\mathcal{D}$ and $\widetilde{f}_{\mathbf{I}}=0$ for
$\mathbf{I}\notin\mathcal{D}$. For open systems, for which $\mathcal{D\neq V}%
$, this extends the vector $f_{\mathbf{I}}$ to the entire lattice in a way
that gives the correct result for this sum. It is thus immediately clear that
$f_{\mathbf{I}}=1$ is a zero eigenvector of $g^{\mathbf{IJ}}$ for closed
systems which is a natural consequence of the translational invariance of
periodic boundaries. For open systems, when $\mathbf{I\pm}2\hat{\mathbf{e}%
}_{\left(  a\right)  }$ falls on the boundary, $\widetilde{f}_{\mathbf{I+}%
2\hat{\mathbf{e}}_{\left(  a\right)  }}=0$ and so the constant vector is no
longer an eigenvector, reflecting the breaking of translational invariance.
However, in both cases, there are other eigenvectors with eigenvalue zero. To
see this, consider one element of the sum on the right hand side of
Eq.(\ref{I1}), say $a=x$. Then the other components of the index are the same
in all terms so, suppressing, those, we see that:%
\begin{align}
v_{1} &  =-\frac{1}{4\Delta^{2}}\left\{  \rho_{2}\left(  \widetilde{f}%
_{3}-\widetilde{f}_{1}\right)  -\rho_{0}\left(  \widetilde{f}_{1}%
-\widetilde{f}_{N_{x}-1}\right)  \right\}  +...\\
v_{2} &  =-\frac{1}{4\Delta^{2}}\left\{  \rho_{3}\left(  \widetilde{f}%
_{4}-\widetilde{f}_{2}\right)  -\rho_{1}\left(  \widetilde{f}_{2}\right)
\right\}  +...\nonumber\\
v_{3} &  =-\frac{1}{4\Delta^{2}}\left\{  \rho_{4}\left(  \widetilde{f}%
_{5}-\widetilde{f}_{3}\right)  -\rho_{2}\left(  \widetilde{f}_{3}%
-\widetilde{f}_{1}\right)  \right\}  +...\nonumber\\
&  ...\nonumber\\
v_{N_{x}-2} &  =-\frac{1}{4\Delta^{2}}\left\{  \rho_{N_{x}-1}\left(
-\widetilde{f}_{N_{x}-2}\right)  -\rho_{N_{x}-3}\left(  \widetilde{f}%
_{N_{x}-2}-\widetilde{f}_{N_{x}-4}\right)  \right\}  +...\nonumber\\
v_{N_{x}-1} &  =-\frac{1}{4\Delta^{2}}\left\{  \rho_{0}\left(  \widetilde
{f}_{1}-\widetilde{f}_{N_{x}-1}\right)  -\rho_{N_{x}-2}\left(  \widetilde
{f}_{N_{x}-1}-\widetilde{f}_{N_{x}-3}\right)  \right\}  +...\nonumber
\end{align}
where the ellipses in the horizontal direction indicate the remaining terms
from the sum over directions , $a=y$ and $a=z$. On the one hand, this clearly
shows that the presence of the boundaries eliminates the zero-mode in which
$f_{\mathbf{I}}=1$ for all $\mathbf{I}$ due to the missing $f_{0}$ terms in
$v_{2}$ and $v_{N_{x}-2}$. In other words, the boundary conditions break the translational symmetry of the differential operators. 
On the other hand, if $N_{x}$ is even, then the ``missing'' elements only occur in 
 even-numbered components of $f$, so they can be avoided by
setting all even-numbered components of $f$ to zero while keeping the
odd-numbered components equal to a constant (e.g. 1). Thus, a spurious zero-mode
exists and its effect on the eigenvectors is clear in the examples below.

It follows from all of this that for a closed system,
$F_{\mathbf{I}}=\lambda\Delta^{3}$ is a stationary state for any constant
$\lambda$ but for open systems this only holds if $\lambda=0$. In order to preserve the correspondence between stationary points on the free energy surface
and stationary states of the dynamics, consider the effect of applying an external one-body
potential $\phi\left(  \mathbf{r}\right)  $ on the system, which modifies the
free energy as $F\left[  \rho;\phi\right]  =F\left[  \rho\right]  +\int
_{V}\rho\left(  \mathbf{r}\right)  \phi\left(  \mathbf{r}\right)
\;d\mathbf{r}$ and so $F\left(  \rho\right)  \rightarrow F\left(  \rho\right)
+\sum_{\mathbf{I\in}\mathcal{D}}\rho_{\mathbf{I}}\phi_{\mathbf{I}}\Delta^{3}$.
Then, the forces become $F_{\mathbf{I}}\left(  \rho,\phi\right)
=F_{\mathbf{I}}\left(  \rho\right)  +\phi_{\mathbf{I}}\Delta^{3}$ so that for
the particular choice $\phi_{\mathbf{I}}=-\lambda$, a stationary state
$F_{\mathbf{I}}\left(  \rho\right)  =\lambda\Delta^{3}$ corresponds to
$F_{\mathbf{I}}\left(  \rho,\phi\right)  =0$ giving a stationary state for
both open and closed systems. Of course, $F\left[  \rho,\phi\right]  =F\left[
\rho\right]  -\lambda\int_{V}\rho\left(  \mathbf{r}\right)  d\mathbf{r}$ is
recognized as the grand-canonical free energy and so one could say that in
order to preserve the correspondence between stationary states in the open and
closed cases of finite systems, one is led naturally to replace the canonical
free energy $F\left[  \rho\right]  $ by the grand canonical $\Omega\left[
\rho;\lambda\right]  =F\left[  \rho\right]  -\lambda\int_{V}\rho\left(
\mathbf{r}\right)  d\mathbf{r}$ when moving from the closed to the open system.

Ito-Stratonovich equivalence can be directly verified with the simple
calculation
\begin{align}
\left(  \frac{\partial}{\partial\rho_{\mathbf{J}}}q_{\mathbf{I}\left(
\mathbf{K}a\right)  }(\rho)\right)  q_{\mathbf{J}\left(  \mathbf{K}a\right)
}(\rho)  &  =\frac{1}{4\Delta^{2}}\sum_{\mathbf{J}\in\mathcal{D}}\sum
_{a\in\left\{  x,y,z\right\}  }\sum_{\mathbf{K\in}\mathcal{V}}\left(
\delta_{\mathbf{I+\hat{\mathbf{e}}_{\left(  a\right)  }K}}-\delta
_{\mathbf{I-\hat{\mathbf{e}}_{\left(  a\right)  }K}}\right)  \frac
{\partial\sqrt{\rho_{\mathbf{K}}}}{\partial\rho_{\mathbf{J}}}\left(
\delta_{\mathbf{J+\hat{\mathbf{e}}_{\left(  a\right)  }K}}-\delta
_{\mathbf{J-\hat{\mathbf{e}}_{\left(  a\right)  }K}}\right)  \sqrt
{\rho_{\mathbf{K}}},\;\;\mathbf{I}\in\mathcal{D}\\
&  =\frac{1}{8\Delta^{2}}\sum_{\mathbf{J}\in\mathcal{D}}\sum_{a\in\left\{
x,y,z\right\}  }\sum_{\mathbf{K\in}\mathcal{V}}\left(  \delta_{\mathbf{I+\hat
{\mathbf{e}}_{\left(  a\right)  }K}}-\delta_{\mathbf{I-\hat{\mathbf{e}%
}_{\left(  a\right)  }K}}\right)  \delta_{\mathbf{JK}}\left(  \delta
_{\mathbf{J+\hat{\mathbf{e}}_{\left(  a\right)  }K}}-\delta_{\mathbf{J-\hat
{\mathbf{e}}_{\left(  a\right)  }K}}\right) \nonumber\\
&  =\frac{1}{8\Delta^{2}}\sum_{\mathbf{J}\in\mathcal{D}}\sum_{a\in\left\{
x,y,z\right\}  }\left(  \delta_{\mathbf{I+\hat{\mathbf{e}}_{\left(  a\right)
}J}}-\delta_{\mathbf{I-\hat{\mathbf{e}}_{\left(  a\right)  }J}}\right)
\left(  \delta_{\mathbf{J+\hat{\mathbf{e}}_{\left(  a\right)  }J}}%
-\delta_{\mathbf{J-\hat{\mathbf{e}}_{\left(  a\right)  }J}}\right) \nonumber\\
&  =0\nonumber
\end{align}
so the model is indeed Ito-Stratonovich equivalent. Furthermore, in exactly
the same way, one finds that
\begin{equation}
\sum_{\mathbf{J} \in \mathcal{D}}q_{\mathbf{I}\left(  \mathbf{K}a\right)  }(\rho)\frac{\partial}{\partial
\rho_{\mathbf{J}}}q_{\mathbf{J}\left(  \mathbf{K}a\right)  }(\rho)=0
\end{equation}
so Stratonovich-anti-Ito equivalence also holds. Thus, all interpretations of
the stochastic differential equation are equivalent.

\subsection{Forward differences scheme A (FWD-A)}

One obvious drawback of the central-differences scheme is the presence of the spurious zero-modes discussed above which breaks the correspondence between stationary states of the dynamics and the free energy.
An alternative is the use of forward differences
such as $\nabla_{a}\rightarrow\frac{1}{\Delta}\left(  \delta_{\mathbf{K}\mathbf{I+}\hat{\mathbf{e}}_{\left(  a\right)  }}-\delta_{\mathbf{K}{\mathbf{I}}}\right)  $ so that
\begin{equation}
q_{a\mathbf{IK}}(\rho)=\frac{1}{\Delta}\left(  \delta_{\mathbf{K}\mathbf{I+}\hat{\mathbf{e}}_{\left(  a\right)  }}-\delta_{\mathbf{K}\mathbf{I}}\right)  \sqrt{\rho_{\mathbf{K}}},\;\mathbf{I}\in\mathcal{D} ,\mathbf{K}\in\mathcal{V}%
\end{equation}
and
\begin{equation}
g_{\mathbf{IJ}}\left(  \rho\right)  =-\frac{1}{\Delta^{2}}\sum_{a\in
\{x,y,z\}}\left(  \rho_{\mathbf{I+\hat{\mathbf{e}}_{\left(  a\right)  }}%
}\left(  \delta_{\mathbf{I+\hat{\mathbf{e}}_{\left(  a\right)  }J}}%
-\delta_{\mathbf{IJ}}\right)  -\rho_{\mathbf{I}}\left(  \delta_{\mathbf{IJ}%
}-\delta_{\mathbf{I-\hat{\mathbf{e}}_{\left(  a\right)  }J}}\right)  \right)
,\;\mathbf{I},\mathbf{J}\in\mathcal{D}%
\end{equation}
It is again easy to show that away from the boundaries
\begin{equation}
\sum_{\mathbf{J}\in\mathcal{D}}g_{\mathbf{IJ}}\left(  \rho\right)
f_{\mathbf{J}}\underset{\Delta\rightarrow0}{\rightarrow}-\left(  \nabla
\cdot\rho\left(  \mathbf{r}\right)  \nabla f\left(  \mathbf{r}\right)
\right)  _{\mathbf{r}_{\mathbf{I}}}.
\end{equation}
For a general vector $f_{\mathbf{J}}$, one has that
\begin{equation}
\sum_{\mathbf{J}\in\mathcal{D}}g_{\mathbf{IJ}}\left(  \rho\right)
f_{\mathbf{J}}=-\frac{1}{\Delta^{2}}\delta_{\mathbf{I}\pm\mathbf{\hat
{\mathbf{e}}_{\left(  a\right)  }}\in\mathcal{D}}\sum_{a\in\{x,y,z\}}\left(
\rho_{\mathbf{I+\hat{\mathbf{e}}_{\left(  a\right)  }}}\left(  \widetilde
{f}_{\mathbf{I+\hat{\mathbf{e}}_{\left(  a\right)  }}}-\widetilde
{f}_{\mathbf{I}}\right)  -\rho_{\mathbf{I}}\left(  \widetilde{f}_{\mathbf{I}%
}-\widetilde{f}_{\mathbf{I-\hat{\mathbf{e}}_{\left(  a\right)  }}}\right)
\right)
\end{equation}
so that again, the constant vector $f_{\mathbf{J}}=1$ is a zero-eigenvector
for the closed case but not for the open case while the spurious
zero-eigenvector that was present for the open system with central differences
is eliminated. As to stationary states, the same conclusions as for central
differences apply:\ $F_{\mathbf{I}}=\lambda$ is a stationary state for closed
systems but not for open systems whereas $\Omega_{\mathbf{I}}=0$ is always stationary.
Checking the Ito-Stratonovich equivalence condition now gives%
\begin{equation}
\sum_{\mathbf{J}\in\mathcal{D}}\sum_{a\in\{x,y,z\}}\sum_{\mathbf{K}%
\in\mathcal{V}}\frac{\partial q_{a\mathbf{IK}}\left(  \rho\right)
}{\partial\rho_{\mathbf{J}}}q_{a\mathbf{JK}}\left(  \rho\right)  =\frac
{1}{2\Delta^{2}}\left(  \delta_{I_{x}N_{x}-1}+\delta_{I_{y}N_{y}-1}%
+\delta_{I_{z}N_{z}-1}\right)
\end{equation}
and so the two interpretations are no longer equivalent, although the
violations are restricted to the boundary.

\subsection{Forward differences scheme B (FWD-B)}

The interpretation-dependence of the stochastic model when formulated with
forward differences can be removed by using a somewhat un-natural formulation
of
\begin{equation}
q_{\mathbf{I}\left(  \mathbf{K}a\right)  }\left(  \rho\right)  =\frac
{1}{\Delta}\left(  \delta_{\mathbf{I+\hat{\mathbf{e}}_{\left(  a\right)  }K}%
}-\delta_{\mathbf{IK}}\right)  \sqrt{\rho_{\mathbf{K+\hat{\mathbf{e}}_{\left(
a\right)  }}}},\;\mathbf{I}\in\mathcal{D},\mathbf{K}\in\mathcal{V}%
\end{equation}
giving
\begin{equation}
g_{\mathbf{IJ}}\left(  \rho\right)  =-\frac{1}{\Delta^{2}}\sum_{a\in
\{x,y,z\}}\left(  \rho_{\mathbf{I}+2\mathbf{\hat{\mathbf{e}}_{\left(
a\right)  }}}\left(  \delta_{\mathbf{I+\hat{\mathbf{e}}_{\left(  a\right)  }%
J}}-\delta_{\mathbf{IJ}}\right)  -\rho_{\mathbf{I+\hat{\mathbf{e}}_{\left(
a\right)  }}}\left(  \delta_{\mathbf{IJ}}-\delta_{\mathbf{I}-\mathbf{\hat
{\mathbf{e}}_{\left(  a\right)  }J}}\right)  \right)  ,\;\mathbf{I,J}%
\in\mathcal{D}%
\end{equation}
and, despite the unnatural form of the differencing scheme, one can verify
that%
\begin{equation}
\lim_{\Delta\rightarrow0}\left(  \sum_{\mathbf{J}\in\mathcal{D}}%
g_{\mathbf{IJ}}f_{\mathbf{J}}\right)  =-\left.  \nabla\cdot\rho\left(
\mathbf{r}\right)  \nabla f\left(  \mathbf{r}\right)  \right\vert
_{\mathbf{r}_{\mathbf{I}}}.
\end{equation}
Checking the Ito-Stratonovich condition one now finds that the anomolous force
vanishes and so the interpretations are indeed equivalent. Similarly, the shifted (i.e. grand-canonical) free energy again preserves the correspondence
between stationary states of the dynamics and the free energy and there are again no spurious zero modes. 

\section{Calculations}

A classic problem for which this model has proven central is the nucleation of
first order phase transitions where understanding the dynamics of critical
clusters (i.e. saddle points on the free energy surface) is critical. Here we
will use the specific case of nucleation of liquid droplets from a vapor as an
example and, specifically, on the eigenvalue and eigenvector of the unstable mode which plays a key
role in calculating the nucleation rate and in identifying the order parameter\cite{Lutsko_Schoonen_JCP}.
Besides a comparison of the effect of the different discretization schemes, we will also use the opportunity
to look at the dependence of the calculated values on the lattice spacing and the size of the computational cell. For the free energy, the standard cDFT model of a hard-core
contribution (based on the explicitly-stable Fundamental Measure Theory)
supplemented with a mean-field contribution is used. Computational details
have been described elsewhere\cite{esFMT, Schoonen, LutskoLam} and will not be
repeated here except to note that we use the \ the Lennard-Jones (LJ)\
interaction potential,
\begin{equation}
v_{LJ}(r)=4\varepsilon\left(  \left(  \frac{\sigma}{r}\right)  _{12}-\left(
\frac{\sigma}{r}\right)  ^{6}\right)
\end{equation}
in our calculations, modified by a  cutoff at some distance $r_{c}$ and a
correspinding shift so that $v\left(  r\right)  -v\left(  r_{c}\right)  $ for
$r<r_{c}$ and zero for larger distances.

\subsection{Unstable Eigenvalues}

As a test of the various implementations, we will use the calculation of the
unstable mode for a stationary state, $\rho^{\ast}$, so that $F_{\mathbf{I}%
}\left(  \rho^{\ast}\right)  =\mu$ for all $\mathbf{I}\in\mathcal{V}$. Then,
writing $\rho_{t\mathbf{I}}=$ $\rho_{\mathbf{I}}^{\ast}+\delta\rho
_{t\mathbf{I}}$ and linearizing, one finds%
\begin{equation}
\frac{\partial\delta\rho_{t\mathbf{I}}}{\partial t}=-\frac{D}{\sigma^{2}}
\sum_{\mathbf{K}\in\mathcal{D}}D_{\mathbf{IK}}\delta\rho_{t\mathbf{K}}%
+\sqrt{\frac{2D}{\Delta^{3}}}\sum_{a\in(x,y,z)}\sum_{\mathbf{J}\in\mathcal{V}%
}q_{\mathbf{I}\mathbf{J}(a)}\left(  \rho^{\ast}\right)  \xi_{t\mathbf{J}%
(a)},\;\mathbf{I}\in\mathcal{D}%
\end{equation}
where the dimensionless \emph{dynamical matrix} is
\begin{equation}
D_{\mathbf{IK}}\equiv\frac{\sigma^{2}}{\Delta^{3}}\sum_{\mathbf{J}%
\in\mathcal{D}}g_{\mathbf{IJ}}\left(  \rho^{\ast}\right)  \beta F_{\mathbf{JK}%
}\left(  \rho^{\ast};\lambda\right)  .
\end{equation}
In the case that $\rho^{\ast}$ is a critical cluster (i.e. a saddle point),
there is at least one unstable mode and the eigenvalue associated with this
mode is a key quantity in calculating the rate of nucleation.

The Hessian of the free energy, $F_{\mathbf{IJ}}$ is a real, symmetric matrix
and so its eigenvalues are all real and its left and right eigenvectors are
the same. The dynamical matrix is not symmetric and so its left and right
eigenvectors are different and if the left eigenvector for the eigenvalue
$\lambda^{(a)}$ is $u^{(a)}_{\mathbf{I}}$ then $v^{(a)}_{\mathbf{I}}
\equiv\sum_{\mathbf{J}\in\mathcal{D}}g_{\mathbf{IJ}}u^{(a)}_{\mathbf{J}}$ is a
right eigenvector with the same eigenvalue. It is straightforward to show that
the eigenvalues are again all real (due to the fact that the dynamical matrix
is the product of two real, symmetric matrices) and that the left and right
eigenvectors form a biorthogonal set. For this reason, it is natural to
normalize them so that $\sum_{\mathbf{J}\in\mathcal{D}}u^{(a)}_{\mathbf{I}%
}v^{(a)}_{\mathbf{I}}=1$ making them bi-orthonormal.

\begin{figure}
[t]\includegraphics[width=0.45\linewidth]{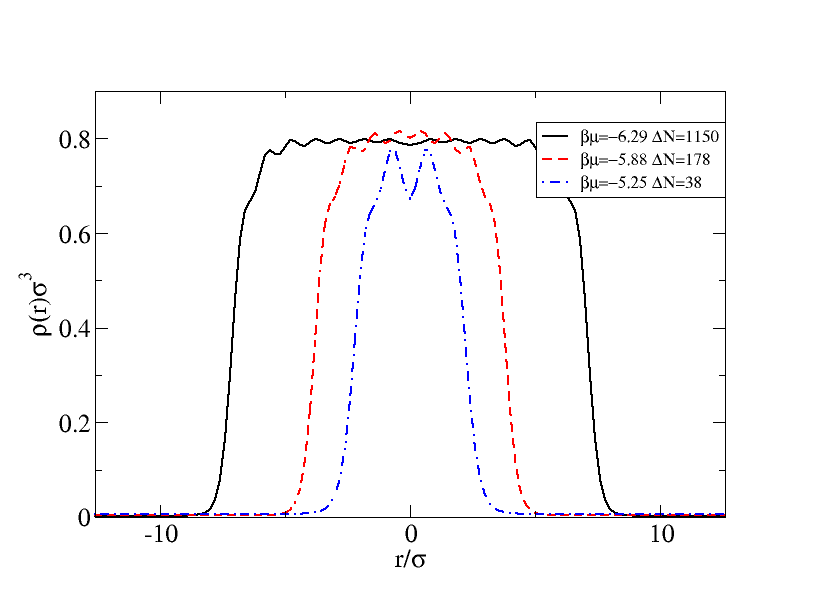}
\includegraphics[width=0.45\linewidth]{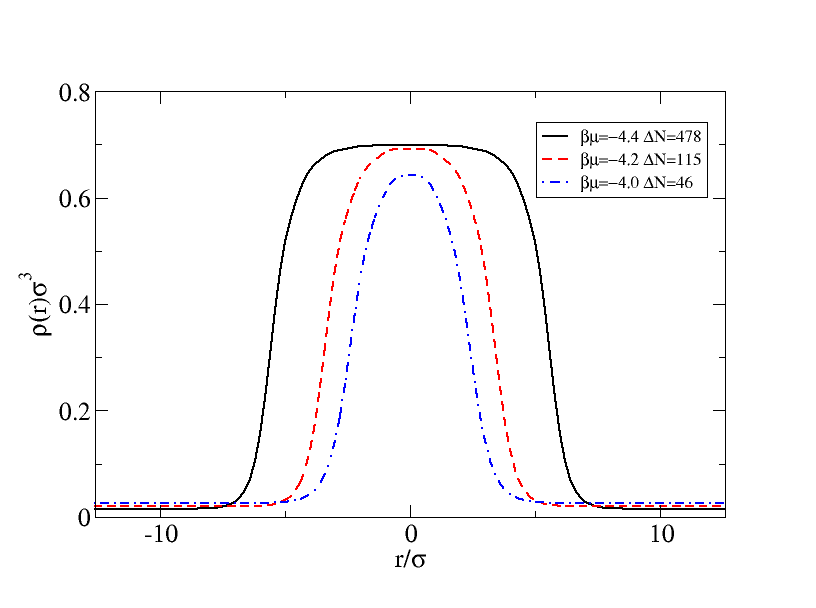}
\caption{The panel on the left shows critical clusters for $k_BT=0.6\varepsilon$ and the panel on the right shows $k_BT=0.8\varepsilon$, both with $\Delta = 0.2\sigma$.}
\label{clusters}
\end{figure}

In order to estimate the nucleation rate\cite{Lutsko_Schoonen_JCP}, it is necessary to determine the
unstable eigenvalues of the dynamical matrix. For the systems considered here,
the dimension of the space of the discretized density space ranges from
$32^{3}$ to $256^{3}$ or from about $3\times10^{4}$ to more than
$1\times10^{7}$, with a corresponding number of eigenvalues. The dynamical
matrix is generally far too large to compute or to hold in memory and
diagonalization is infeasible. It is, however, possible to determine a
restricted set of eigenvalues and eigenvectors. We do this using the SLEPC
library\cite{SLEPC1,SLEPC2,SLEPC3,SLEPC4}. After experimenting with many
different routines, we find good results using the Jacobi-Davidson solver
(EPSJD) when determining eigenpairs for the Hessian of the free energy and the
Krylov-Schur solver (EPSKRYLOVSCHUR) (a variation of the Arnoldi method) for
the dynamical matrix. The SLEPC library is dependent on the PETSc
linear-algebra library and in both cases, the Generalized Minimal Residual
method is used for solving linear systems (KSP\ type KSPGMRES in PETSc).

Critical clusters were created as described above based either on the correspondence
between stationary states for open and closed systems or on the force method.
Figure \ref{clusters} shows some typical examples of the critical clusters
obtained for a cubic computational cell with sides of length $L=25.6\sigma$. We recall that for a cutoff of $R_{c}=3\sigma$, the liquid-vapor
critical temperature determined from the cDFT model is approximately
$k_{B}T_{c}=1.295\varepsilon$ while the triple point is $k_{B}T_{c}%
=0.798\varepsilon$.

Tables \ref{tab1} and \ref{tab2} summarize calculations for a critical droplet
in an open system using different values of the computational lattice
parameter and the different discretization schemes discussed. The results
using the two forward-difference schemes are in all cases very similar and are
relatively independent of the lattice spacing for $\Delta\lesssim0.4\sigma$.
Physically, this is consistent with the smallest features of the density
profiles being oscillations of wavelength around $\sigma$ in the cooler system
and gradients $d\rho\sigma^{3}/d\left(  r/\sigma\right)  \sim0.3-0.5$ in the
interfacial regions giving relative density variations over one computational
cell of $\Delta\times d\ln\rho/dr\lesssim0.4$ at the middle of the interfacial
region. On the other hand, some variation is expected as the global properties
of the clusters, such as their total excess free energy and mass, vary
slightly with lattice spacing as seen in the tables. However, the
central-difference results differ substantially from the forward-differences
as well as showing much stronger lattice-spacing dependence, particularly at
the higher temperature.

\begin{table}
[ptb]\begin{minipage}[v]{\linewidth} 
\addtolength\tabcolsep{2pt}
\begin{tabular}[c]{|*3c|*3c|}%
\hline
& & & \multicolumn{3}{|c|} {$|\lambda| \times 10^3$} \\
\hline			
$\Delta$ & $\beta \Delta \Omega$ & $\Delta N$ & FWD-A  & FWD-B & Central \\
\hline
$0.1$ & $62.7$ & $260$ & $1.09$ & $1.09$ & $1.05$\\
$0.2$ & $62.3$ & $257$ & $1.10$ & $1.10$ & $0.52$\\
$0.4$ & $61.1$ & $251$ & $1.12$ & $1.14$ & $0.54$\\
$0.8$ & $69.0$ & $314$ & $0.85$ & $0.85$ & $0.22$\\
\hline
\end{tabular}
\caption{Sensitivity of the excess free energy, cluster mass and unstable eigenvalue to the calculational lattice spacing for $k_BT = 0.8 \varepsilon$ and $\beta \mu = -4.61$. The potential cutoff was $R_c=5\sigma$ and the volume of the computational cell was held fixed at $V=\left(25.6\sigma\right)^3$ in all calculations.}
\label{tab1}
\end{minipage}

\end{table}

\begin{table}
[ptb]\begin{minipage}[v]{\linewidth} 
\addtolength\tabcolsep{2pt}
\begin{tabular}[c]{|*3c|*3c|}%
  \hline
& & & \multicolumn{3}{|c|} {$|\lambda| \times 10^3$} \\  
\hline			
$\Delta$ & $\beta \Delta \Omega$ & $\Delta N$ & FWD-A  & FWD-B & Central \\
\hline
$0.1$ & $21.6$ & $38.3$ & $5.26$ & $5.27$ & $4.66$\\
$0.2$ & $21.5$ & $38.0$ & $5.30$ & $5.33$ & $4.69$ \\
$0.4$ & $21.2$ & $37.2$ & $5.43$ & $5.55$ & $4.81$\\
$0.8$ & $21.5$ & $38.0$ & $4.31$ & $4.59$ & $3.77$\\
\hline
\end{tabular}
\caption{Sensitivity of the excess free energy, cluster mass and unstable eigenvalue to the calculational lattice spacing for $k_BT = 0.6 \varepsilon$ and $\beta \mu = -5.25$. The potential cutoff was $R_c=3\sigma$ and the volume of the computational cell was held fixed at $V=\left(25.6\sigma\right)^3$ in all calculations.}
\label{tab2}
\end{minipage}

\end{table}

\begin{figure}[t]%
\includegraphics[scale=0.3]{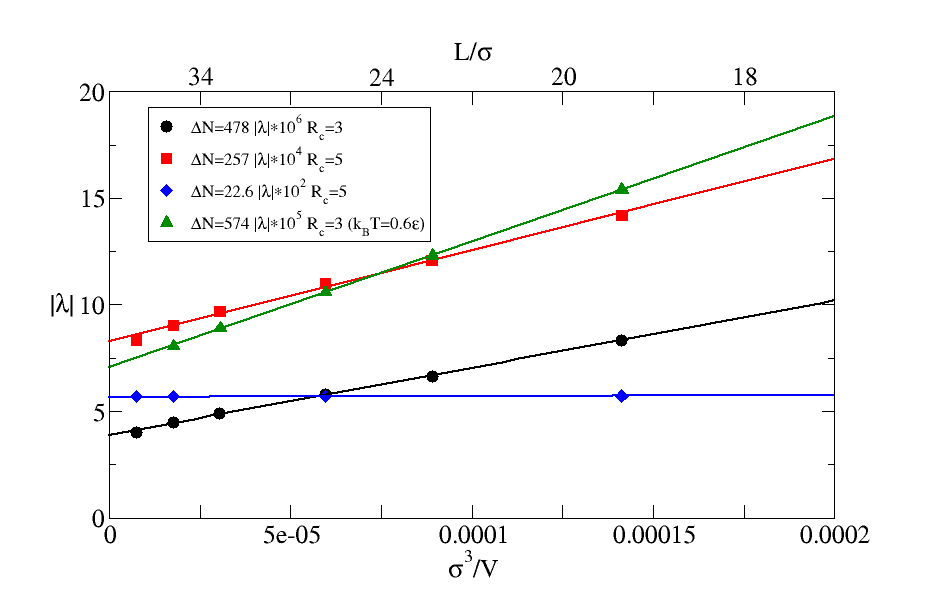}
\caption{Unstable eigenvalues as functions of the inverse system size as calculated using the {\bf FWD-A} scheme. The legend gives, for each curve, the size of the critical cluster ($\Delta N$), the scaling of the magnitude of the dimensionless (unstable) eigenvalues $\lambda$ (done for display purposes) and the potential cutoff used in each calculation. All calculations were performed at $k_BT=0.8\varepsilon$ except that shown as green triangles which, as noted in the legend, was done for $k_BT=0.6\varepsilon$. The scale below the graph shows the abscissa in terms of the volume of the cubic computational cell and that above the graph shows it as the length of the computational cell's sides.}
\label{sys_size}
\end{figure}

In the open system, the boundaries act as sources and sinks of density whereas inside the system, the movement of mass is conservative and diffusive. This naturally raises the question
of the sensitivity of the calculations to system size (for fixed computational lattice spacing). This is illustrated in Figure \ref{sys_size} which shows the value of the unstable eigenvalues as functions of the system size for lattice spacing $\Delta = 0.2\sigma$. For the smallest cluster considered, containing only about 23 molecules, the eigenvalue is virtually independent of box size. However, for the larger clusters, the results seem to scale with the inverse volume of the system but only vary by about a factor of 2 over the range of available box sizes (the boxes have a minimum size in order to contain the cluster and a maximum size based on computational feasibility). To give some context to the observed behaviour, we note that the clusters having excess masses of $\Delta N = 22.7, 257, 478$ and $573$ have radii (based on a fit to a hyperbolic tangent profile\cite{LutskoBubble1}) of $3.2\sigma, 8.8\sigma, 11\sigma$ and $11.2\sigma$, respectively. 

\subsection{Unstable eigenvectors}

\begin{figure}
[t]\includegraphics[width=0.45\linewidth]{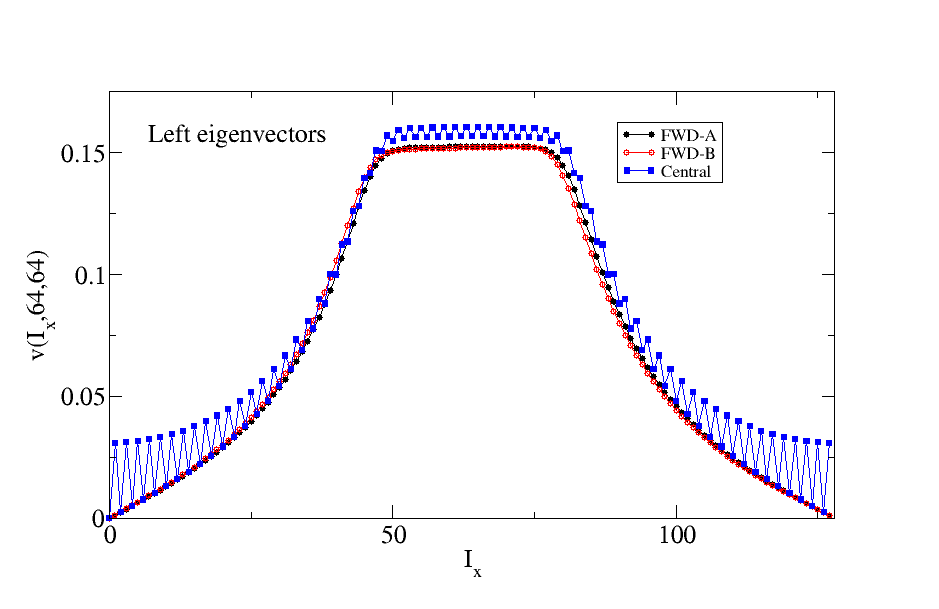}
\includegraphics[width=0.45\linewidth]{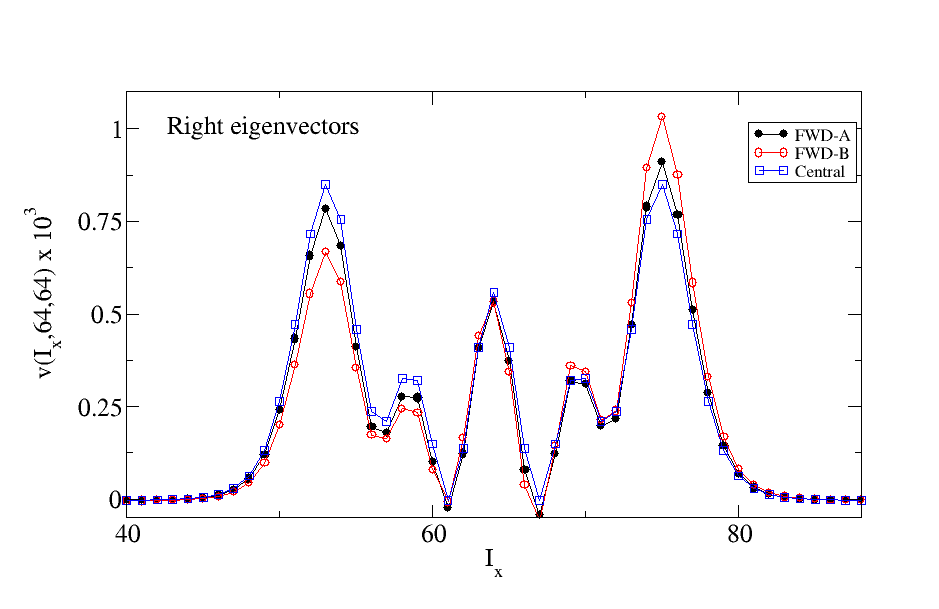}
\caption{Left and right eigenvectors for $k_BT = 0.6\varepsilon$,  $\beta \mu = -5.25$ and $\Delta = 0.2\sigma$ for the three schemes. The computational cell has $128$ lattice points in each direction and the figure shows a slice through the volume that characterizes the (approximately) spherically symmetric eigenvectors.}
\label{eigen}
\end{figure}

\begin{figure}[t]%
\includegraphics[width=0.45\linewidth]{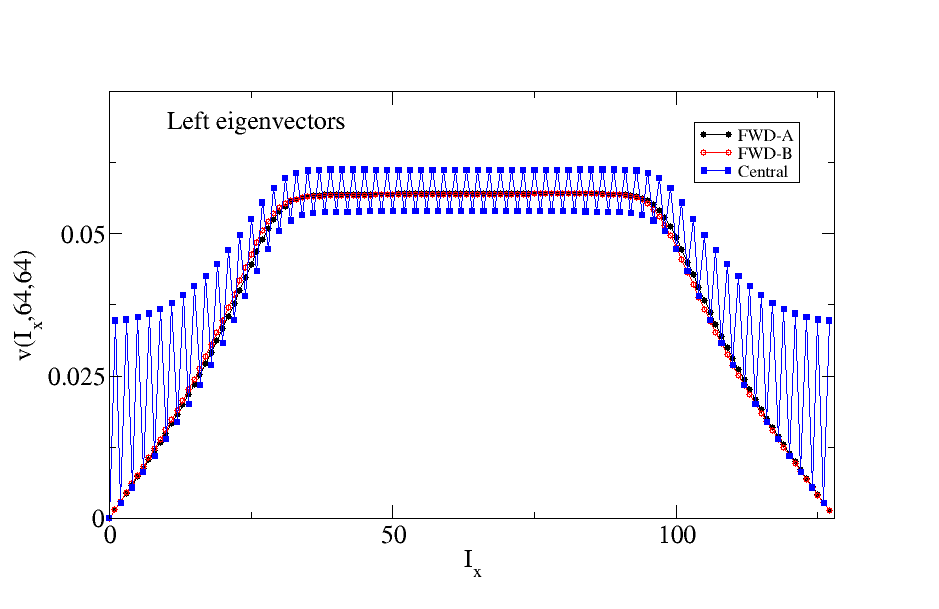}
\includegraphics[width=0.45\linewidth]{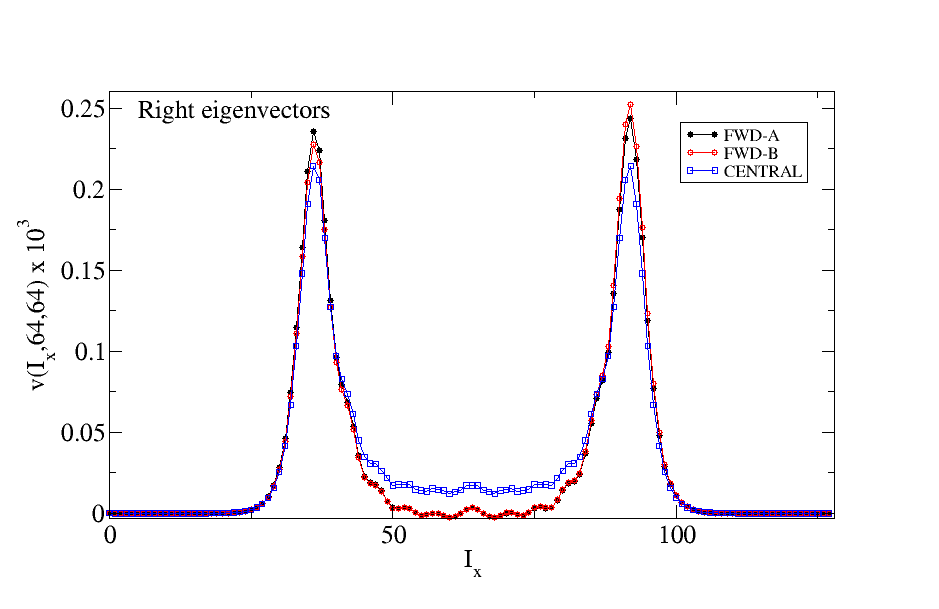}
\caption{Left and right eigenvectors for $k_BT = 0.8\varepsilon$,  $\beta \mu = -4.4$ and $\Delta = 0.2\sigma$ for the three schemes. The computational cell has $128$ lattice points in each direction and the figure shows a slice through the volume that characterizes the (approximately) spherically symmetric eigenvectors.}
\label{eigen2}
\end{figure}

To understand the differences between the FWD and the CENTRAL\ scemes, it is
useful to consider the corresponding eigenvectors. Fig.\ref{eigen} shows the
left and right eigenvectors of the unstable mode for a small cluster at
$k_{B}T=0.6\varepsilon$. The right eigenvectors are quite similar for all
three discretization schemes with a notable difference being that the
eigenvectors for the forward schemes show some asymmetry which is presumeably
an echo of the un-symmetric nature of the difference scheme. The
central-differences scheme is, in contrast, highly symmetric. The left
eigenvectors tell a very different story: the forward-difference schemes give
smooth curves which are very similar for the two schemes. In contrast,
the central difference scheme shows strong oscillations on the scale of the
computational lattice, presumeably attributable to the partial decoupling of
the even and odd sublattices as discussed above. These results are typical and
for comparison, Fig. \ref{eigen2} shows a case at $k_{B}T=0.8\varepsilon$
which is qualitatively the same.

\section{Conclusions}

Our results show that the Dean-Kawasaki type model, at least as used in the
description of nucleation, can be modified to account for open boundaries by
fixing the value of the dynamical variables on the boundary. However, this
raises various secondary issues which must be accounted for, namely:

\begin{itemize}
\item The free energy that drives the dynamics must be changed from a
canonical form, applicable to systems with fixed number of particles, to a
grand-canonical form in order to preserve the stationary states whihc are
expected to be the same in both cases.

\item The fixed boudaries mean that Ito-Stratonovich equivalence is sometimes
violated at the boundaries, depending on the discretization scheme. Central
differences and specially constructed forward-difference schems maintain the
equivalence but some natural choices do not.

\item Central differences give a partial decoupling of the even and odd
lattices with the result that a spurious zero-eigenvalue mode appears. This is
avoided with the forward schemes.
\end{itemize}

As an illustration, the various schemes described here were used to calculate
the unstable eigenvalue for a droplet nucleating from a vapor phase. The
forward schemes showed relatively weak dependence on the lattice spacing while
the central difference scheme was much more sensitive. Furthermore, the left
eigenvectors of the central scheme showed spurious oscillations on the scale
of the computational lattice spacing that one assumes is due to the decoupling
of the lattices. Taken together, these results suggest that either of the forward difference
schemes could be used in practical calculations but the specially
constructed FWD2 has the best formal properties of the three schemes examined. 

Finally, the present analysis was performed for the Kawasaki-Dean dynamics
that results from the over-damping of the full fluctuating hydrodynamics
model. The present results suggest that it should be possible to formulate an
open-system version of the latter along similar lines although a detailed
analysis would be necessary to be sure.

\begin{acknowledgments}
This work was supported by the European Space Agency (ESA) and the Belgian
Federal Science Policy Office (BELSPO) in the framework of the PRODEX
Programme, Contract No. ESA AO-2004-070.
\end{acknowledgments}

\bibliography{./mnt_1.bib}

\begin{appendix}
  
\section*{Fluctuation-Dissipation relation}\label{FlucDiss}
Recall that
\begin{align}
g(\mathbf{r},\mathbf{r}';[\rho_t]) &= -\nabla \cdot \rho_t(\mathbf{r}) \nabla \delta(\mathbf{r}-\mathbf{r}')  \\
q_a(\mathbf{r},\mathbf{r}';[\rho_t]) &= \nabla_a \sqrt{2 \rho_t(\mathbf{r})} \delta(\mathbf{r}-\mathbf{r}'),  \nonumber
\end{align}
so
\begin{align}
\int_V \sum_{a \in \{x,y,z\}}q_a(\mathbf{r},\mathbf{r}';[\rho_t])q_a(\mathbf{r''},\mathbf{r}';[\rho_t]) d\mathbf{r}' & =  \sum_{a \in \{x,y,z\}} \int_V \left\{\nabla_a \sqrt{2 \rho_t(\mathbf{r})} \delta(\mathbf{r}-\mathbf{r}')\right\}\left\{\nabla''_a \sqrt{2 \rho_t(\mathbf{r''})} \delta(\mathbf{r''}-\mathbf{r}')\right\} d\mathbf{r}' \\
& =  \sum_{a \in \{x,y,z\}} \nabla_a \sqrt{2 \rho_t(\mathbf{r})}\nabla''_a \sqrt{2 \rho_t(\mathbf{r''})}\int_V  \delta(\mathbf{r}-\mathbf{r}') \delta(\mathbf{r''}-\mathbf{r}')d\mathbf{r}' \nonumber \nonumber \\
& =  \sum_{a \in \{x,y,z\}} \nabla_a \sqrt{2 \rho_t(\mathbf{r})}\nabla''_a \sqrt{2 \rho_t(\mathbf{r''})}\delta(\mathbf{r}-\mathbf{r}'') \nonumber
\end{align}
To further simplify, consider the action of this operator on an arbitrary test function, $f(\mathbf{r})$,
\begin{align}
\int_V\sum_{a \in \{x,y,z\}} \nabla_a \sqrt{2 \rho_t(\mathbf{r})}\left\{\nabla''_a \sqrt{2 \rho_t(\mathbf{r''})}\delta(\mathbf{r}-\mathbf{r}'')\right\}f(\mathbf{r}'') d\mathbf{r}'' & = \sum_{a \in \{x,y,z\}} \nabla_a \sqrt{2 \rho_t(\mathbf{r})}\int_V\left\{\nabla''_a \sqrt{2 \rho_t(\mathbf{r''})}\delta(\mathbf{r}-\mathbf{r}'')\right\}f(\mathbf{r}'') d\mathbf{r}''
\end{align}
but
\begin{align}
\int_V\left\{\nabla''_a \sqrt{2 \rho_t(\mathbf{r''})}\delta(\mathbf{r}-\mathbf{r}'')\right\}f(\mathbf{r}'') d\mathbf{r}'' & = \int_V\nabla''_a\left\{ \sqrt{2 \rho_t(\mathbf{r''})}\delta(\mathbf{r} -\mathbf{r}'')f(\mathbf{r}'')\right\} d\mathbf{r}'' \\ & -\int_V\sqrt{2 \rho_t(\mathbf{r''})}\delta(\mathbf{r}-\mathbf{r}'')\nabla''_af(\mathbf{r}'') d\mathbf{r}'' \nonumber
\end{align}
The first term on the right vanishes for a rectilinear integration volume with periodic boundaries or an infinite system in which everything becomes uniform at sufficiently large distance from the origin. Then,
\begin{align}
\int_V\left\{\nabla''_a \sqrt{2 \rho_t(\mathbf{r''})}\delta(\mathbf{r}-\mathbf{r}'')\right\}f(\mathbf{r}'') d\mathbf{r}'' & =  -\int_V\sqrt{2 \rho_t(\mathbf{r''})}\delta(\mathbf{r}-\mathbf{r}'')\nabla''_af(\mathbf{r}'') d\mathbf{r}'' \\
& =  -\sqrt{2 \rho_t(\mathbf{r})}\nabla_af(\mathbf{r})   \nonumber
\end{align}
and
\begin{align}
\int_V\sum_{a \in \{x,y,z\}} \nabla_a \sqrt{2 \rho_t(\mathbf{r})}\left\{\nabla''_a \sqrt{2 \rho_t(\mathbf{r''})}\delta(\mathbf{r}-\mathbf{r}'')\right\}f(\mathbf{r}'') d\mathbf{r}'' & = -2\nabla_a  \rho_t(\mathbf{r})\nabla_af(\mathbf{r}) \\
& = -2 \int_V g(\mathbf{r}, \mathbf{r}'';[\rho_t]) f(\mathbf{r}'') d\mathbf{r}'' \nonumber
\end{align}
thus showing that
\begin{equation}
\int_V \sum_{a \in \{x,y,z\}}q_a(\mathbf{r},\mathbf{r}';[\rho_t])q_a(\mathbf{r''},\mathbf{r}';[\rho_t]) d\mathbf{r}' = 2 g(\mathbf{r}, \mathbf{r}'';[\rho_t])
\end{equation}
which is the fluctuation-dissipation relation.
\end{appendix}

\end{document}